\documentclass[journal=jacsat,manuscript=letter]{achemso}
\usepackage{chemformula} 
\usepackage[T1]{fontenc} 
\usepackage{amsmath} 
\usepackage {units}
\usepackage{here}    
\usepackage{amsmath,amssymb}
\usepackage{graphicx}   
\usepackage{verbatim}   
\usepackage{color}      
\usepackage{bm}
\usepackage{subfigure}  
\usepackage{hyperref}   
\usepackage[normalem]{ulem} 
\usepackage{siunitx} 
\usepackage[pagewise]{lineno}

\author{Shunya Araki}
\affiliation[kyushu]{Department of Physics, Kyushu University, Motooka 744, Fukuoka 819-0395, Japan}\altaffiliation{Contributed equally to this work}
\author{Kazusa Beppu}
\affiliation[kyushu]{Department of Physics, Kyushu University, Motooka 744, Fukuoka 819-0395, Japan}\altaffiliation{Contributed equally to this work}
\author{Arif Md. Rashedul Kabir}
\affiliation[hokkaido]{Faculty of Science, Hokkaido University, Kita 10, Nishi 8, Kita-ku, Sapporo 060-0810, Hokkaido, Japan}
\author{Akira Kakugo}\email{kakugo@sci.hokudai.ac.jp}
\affiliation[hokkaido]{Faculty of Science, Hokkaido University, Kita 10, Nishi 8, Kita-ku, Sapporo 060-0810, Hokkaido, Japan}
\author{Yusuke T. Maeda}
\affiliation[kyushu]{Department of Physics, Kyushu University, Motooka 744, Fukuoka 819-0395, Japan}
\email{ymaeda@phys.kyushu-u.ac.jp}

\title{Tailoring collective motion of kinesin-driven microtubules via topographic landscapes}


\begin{document}

\begin{abstract}
Biomolecular motor proteins that generate forces by consuming chemical energy obtained from ATP hydrolysis are pivotal for organizing broad cytoskeletal structures in living cells. The control of such cytoskeletal structures benefits programmable protein patterning; however, our current knowledge is limited owing to the underdevelopment of an engineering approach for controlling pattern formation. Here, we demonstrate the tailoring of assembled patterns of microtubules (MTs) driven by kinesin motors by designing the boundary shape in fabricated microwells. We found an MT bundle structure along the microwell wall and a bridging structure perpendicular to the wall. Corroborated by the theory of self-propelled rods, we further showed that the alignment of MTs defined by the boundary shape determined the transition of the assembled patterns, providing a blueprint to reconstruct bridge structures in microchannels. Our findings provide a geometric rule to tailor the self-organization of cytoskeletons and motor proteins for nanotechnological applications.
\end{abstract}

\section{Introduction}
The self-organization of cytoskeletal proteins into propagating waves or long, bundled structures in living cells controls highly complex events such as cell division, cell migration, and intracellular material transport \cite{needleman}. Ordered cytoskeletal structures are self-organized under the confinement of the cell-sized space, as observed in ring-like cytoskeletons involved in the division of rod-shaped bacteria \cite{schwille}. The understanding of such cytoskeletal structures can elucidate cell mechanics related to subcellular physiology and serve as a foundation for protein-based functional devices. In recent decades, attempts to precisely manipulate enzymatic reactions promoting the self-organization of the cytoskeleton have been achieved by using optical tools \cite{diez} and magnetic nanoparticles \cite{zoher1,zoher2}. Although these techniques are useful in controlling local structures, the use of such narrowly focused external fields alone limits the ability to expand patterning on a global scale (e.g., wide-field laser scanning). In nanotechnological applications, increasing attention is being paid to cytoskeletal assemblies driven by biomolecular motor proteins \cite{heuvel}. 

Biomolecular motor proteins generate forces or torques autonomously by consuming the chemical energy obtained from ATP hydrolysis. Kinesin is a major motor protein that moves unidirectionally along the hollow rod-shaped cytoskeleton composed of tubulin protein, called microtubules (MTs). Motile MTs driven by kinesin are versatile experimental models for studying the self-organization of proteins \cite{dogic1,sumino,senoussi,kakugo1}. When kinesin is bound to its substrate with abundant ATP and MTs at high densities, the motile MTs align their heading angle and show collective motion, similar to vortex lattices, and ordered streaming occurs in the absence of complex enzymatic reactions or manipulation by external forces \cite{kakugo2,kakugo3,yoshinaga,tanida}. 

A promising approach for designing motile MTs with collective motion in nanotechnology applications, which has been overlooked to date, is to elicit geometrical control of the self-organized structure of cytoskeletons \cite{dogic3}. The control of binary collisions of MTs is the key to tailoring new ordered patterns in microfabricated devices \cite{wioland1,beppu1,nishiguchi,reinken,cammann,beppu2}. For instance, when dense MT suspensions are enclosed within a microchannel with kinesin motors, chaotic streaming of a confined fluid solution emerges, which can drive material transport spontaneously \cite{dogic4}. However, these studies are limited to simple channel shapes because the methodology for geometric design that controls MT interactions has not been explored using built-in topography \cite{diez,dekker}. Hence, the topographic engineering of their collective motion offers a novel strategy to construct functional devices powered by motor proteins \cite{kakugo4}. 

Here, we sought to demonstrate the tailoring of collective motion in kinesin-driven active cytoskeletons by corroborating geometric microdevices and theoretical models of motile MTs. The MTs are propelled by the kinesin motor under spatial confinement, and the boundary shape of the microwells is used to precisely guide the binary interactions of motile MTs. We found two ordered patterns of MTs in the designed microwells: protrusion-like bundles emanating from the wall obstacle and bridge formation across the facing boundaries. The transition of MT patterns was determined by a geometric rule involving the angle of binary collision, suitably corresponding to the theoretical model of self-propelled rods. The application of the geometric rule will facilitate the rational design of an MT trail ordered along the microchannel as well as a bridge structure that forms a dividing line across the microchannel.

\section{Results and discussion}

We constructed a microfabricated system in which MTs driven by kinesin molecular motors were enclosed in microwells with a predesigned boundary shape \cite{beppu1,izri,beppu2}. The kinesin used in this study was a recombinant kinesin consisting of the first 573 amino acid residues of human kinesin-1 \cite{kabir}. Rhodamine-labeled MTs were prepared by polymerizing a mixture of rhodamine-labeled tubulin (RT) and non-labeled tubulin (WT) for 30 min at \SI{37}{\degreeCelsius} and a molar ratio of (RT:WT) 4:1. After fabrication, polydimethylsiloxane (PDMS) microwells were attached to a glass slide, and the flow cell was constructed using a double-sided tape (Figure 1(a)). Kinesin proteins were adsorbed onto the PDMS surface, following which MTs were deposited at various densities by injection into the chamber. Inside the microwells, MTs can be attached to the kinesin motors at the surface and spontaneously move by consuming ATP. The length of the MT is 18.9$\pm$\SI{14.3}{\micro\meter}, and its density is 0.88 filaments/$\mu$m$^{2}$ (Figure S1), unless otherwise specified. The self-propelled MTs interact with each other by near-field attraction because of the depletion force of 0.3\% methyl cellulose in the solution and steric interactions among MTs \cite{kakugo1}. The velocity of self-propelled MT is 0.24$\pm$\SI{0.04}{\micro\meter\per\second} in 0.3\% methylcellulose solution.

First, we confined the kinesin-MT active cytoskeleton in a circular microwell with a radius of $R = \SI{82}{\micro\meter}$ and height of $h = \SI{25}{\micro\meter}$ to test the effective confinement of the designed microwells (Figure 1(b)). MTs that collide with the wall also interact with the boundary in a nematic manner, causing them to move along the wall (Figure 1(a)). For the first \SI{10}{\minute}, the MTs moved collectively in the microwell with an aligned nematic orientation; however, after \SI{30}{\minute}, they begin to move along the wall. The bundles of motile MTs were found to be localized close to the boundary wall around \SI{60}{\minute}, suggesting that the confinement steers the collective motion of motile MTs and traps them close to the wall.

To guide the direction of collectively migrating MTs, we designed overlapping circles (doublet and triplet circles) of confined microwells (Figures 1(c) and 1(d)). Because the angle of the binary collision among MTs is important for the geometric control of collective motion, the doublet and triplet circle boundaries are suitable for facilitating the collision of MTs oriented along the boundary \cite{beppu1,beppu2}. For the doublet microwell, the boundary is defined by two overlapping circles with radius $R$ as well as the distance between the centers of two circles $\Delta$ (Figure 1(c)). We examined the ordered patterns of alignment of dense self-propelled MTs at various $\Delta$ distances, from \SI{41}{\micro\meter} to \SI{161}{micro\meter} while maintaining the same radius $R = \SI{82}{\micro\meter}$. In the doublet microwell with a short $\Delta$ distance ($\Delta = \SI{96}{\micro\meter}$, $\Delta/R = 1.17$), the MTs formed a protrusion extending laterally at right angles from the \textit{tip} of the doublet circle (Figure 1(c), top). When the overlapping circles were separated over long $\Delta$ distances ($\Delta = \SI{157}{\micro\meter}$, $\Delta/R = 1.92$), the protrusions from the two tip regions facing each other were fused into a bridge-like structure (Figure 1(c), bottom). 

In addition, the hierarchical patterns created by the connection of bridges or protrusions from the tip are also important for the tailoring of ordered MT assemblies. For this purpose, we employed the triplet circle boundary, in which the tips were arranged in an equilateral triangle, and three circles were superimposed. Figure 1(d) shows MT assembly in the microwells with a triplet circle boundary of $R=\SI{82}{\micro\meter}$. With the distance $\Delta = \SI{96}{\micro\meter}$ (and $\Delta/R = 1.17$), a protrusion pattern emerged from the three tips, similar to the protrusion in the doublet circle boundary (Figure 1(d), top). As the distance $\Delta$ became enlarged ($\Delta = \SI{157}{\micro\meter}$ and $\Delta/R=1.92$), the bridges of the aligned MTs were created from the three tips, and the bridges formed a pattern in which they were oriented toward each other (Figure 1(d), bottom). The collectively aligned MTs, even those with complex geometries, could be easily controlled by the geometric parameter $\Delta/R$.

Because the geometric parameter, $\Delta/R$, is defined by the angle of binary collision, $\Psi$, at the tip of overlapping circles ($\Delta/R = 2\cos\Psi$), the observed active cytoskeletal patterns reflect the geometry-dependent collective motions under nematic alignment at the tip. Next, we quantitatively investigated the transition from the protrusion pattern to bridge formation, by varying $\Delta/R$ from 0.5 to 1.97 while keeping $R=\SI{82}{\micro\meter}$ unchanged. In the doublet circle boundary, the protrusion pattern was found at $\Delta/R<1.33$, whereas the bridge formation occurred at $\Delta/R\geq1.33$ (Figure 2(a)). The transition point was also similar in the triplet circle boundary; the bridges were formed at $\Delta/R>1.33$ (Figure 2(b)). Moreover, we analyzed the orientation field of MTs $\theta(\bm{r})$ at position $\bm{r}$ in doublet and triplet circle boundaries (Figure 2(c) and 2(d), see Supplemental Methods for details). The transition from the protrusion to bridge pattern was examined using the order parameter, $\Phi$, defined as $\Phi=\langle \cos2(\theta(\bm{r}) - \theta_{bdg})\rangle_{ROI}$, where $\theta_{bdg}$ is the angle of the axis from the centroid of the microwell to the tip, and $\langle \cdot \rangle_{ROI}$ denotes the ensemble average of the angle observed from the centroid-to-tip axis (Figure 2(e)). The order parameter, $\Phi$, monotonically increased upon increasing the value of $\Delta/R$ for microwells of different sizes ($R=46, 82, 91$, \SI{100}{\micro\meter}). This geometry-induced transition is shared with the triplet circle boundary upon the application of increasing $\Delta/R$ values (Figure 2(f)). 

Next, we clarified whether the parameter, $\Delta/R$, is a geometric measure for the transition of patterns, even at various concentrations of MTs. Because the density of an actively moving agent is important for the onset of collective motion, we examined the geometric dependence of bridge/protrusion formation over a range of density values, from 0.22 filaments/$\mu$m$^{2}$ (Figure 2(e) and 2(f), deep green), which is the lowest MT density at which collective motion occurs (Figures S1 and S2), to 1.54 filaments/$\mu$m$^{2}$ (Figure 2(e) and 2(f), purple). We set the threshold order parameter of $\Phi_c = 0$ to distinguish between protrusion and bridge patterns and allow the classification of pattern formation at various microwell sizes ($R= 46, 82, 91$, and \SI{100}{\micro\meter}) and different MT densities (0.22, 0.88, 1.54 filaments/$\mu$m$^{2}$). For the doublet circle boundary, the order parameter, $\Phi$, became $\Phi > 0$, and the transition from the protrusion pattern to bridge structure occurred near $\Delta/R=1.3-1.4$ (Figure 2(e) and 2(g), left), indicating that boundary geometry could control the transition from protrusion to bridge pattern over a wide range of MT densities and confinement sizes. In addition, the assembled patterns of MTs in the triplet circle boundary showed the transition of two patterns ($\Phi \approx 0$) near $\Delta/R=1.5-1.6$ (Figure 2(f) and 2(g), Right). This slight shift of the transition point indicated that bridge structures in the triplet circle boundary were connected to each other in a hierarchical pattern and that the orientation angle could be expanded.

We then sought to determine why the pattern formation of self-propelled MTs was precisely regulated by the single geometric parameter $\Delta/R$. To this end, we theoretically analyzed geometry-dependent collective motion using a self-propelled rod model \cite{vicsek,chate1}. We consider rods whose orientation is aligned by a nematic orientation interaction as self-propelled. The position of the rod $m$ at time $t$ is $\bm{r}_m(t)$, and the velocity is $\bm{v}_m(t)=v_0\bm{e}(\theta_m)$ with the unit vector of $\bm{e}(\theta_m)$=$(\cos\theta_m(t),\sin\theta_m(t))$. Given that the orientation interaction between the rods is defined by the potential $U(\bm{r}_m,\theta_m)$, the dynamics of self-propelled rods is given by the following equations: 
\begin{equation}
\dot{\bm{r}}_m(t) = v_0 \bm{e}(\theta_m),
\end{equation}
\begin{equation}
\dot{\theta}_m(t) = -\bar{\gamma} \frac{\partial U}{\partial \theta} + \eta_m(t),
\end{equation}
where $\bar{\gamma}$ is the strength of the nematic alignment of the rods, and $\eta_m(t)$ is the random noise in rod orientation ($\langle \eta_m \rangle = 0$ and $\langle \eta_m(t)\eta_n(t') \rangle = 2D\delta_{mn}\delta(t-t')$ with the angular diffusion coefficient $D$). The potential of the nematic alignment is given by the following equation:
\begin{equation}
U(\bm{r}_m,\theta_m) = - \sum_{|\bm{r}_{mn}|<\epsilon} \cos 2(\theta_m - \theta_n)
\end{equation}
where $|\bm{r}_{mn}| = |\bm{r}_m - \bm{r}_n|$ is the distance between the rods, and $\epsilon$ is the effective radius of nematic alignment. The self-propelled rods move inside the confined space whose boundary shape is defined by overlapping circles, as shown in the experiments (Figures 1 and 2), with an elevation angle of $\Psi$ at the tip (Figure 3(a)). Numerical simulation of self-propelled rods under a doublet circle boundary revealed the transition from the protrusion pattern to bridge pattern, as observed in our experiments (Figure 3(b) top, Movies S1 and S2). The transition of the protrusion to the bridge pattern occurs at $\Delta/R \approx 1.4$, which correspond to the experimental transition points (Figure 2). A similar transition is also observed for self-propelled rods confined under the triplet circle boundary and the same geometric condition $\Delta/R \approx 1.4$ (Figure 3(b) bottom, Movies S3 and S4). 

To further analyze the transition of the protrusion pattern to the bridge, we examined the effective potential that governs rod orientation around the tip of overlapping circular geometries. The nematic alignment of self-propelled rods at this tip can take four orientation angles $\pi/2 \pm \Psi$ and $- \pi/2 \pm \Psi$ (Figure 3(a)). By substituting these orientation angles in Eq. (3), the effective potential yield was $U(\theta;\Psi) = \frac{1}{2}\cos2\theta\cos2\Psi$ (Figure 3(c)). Importantly, at large $\Psi$ values, this potential landscape has local minima at two angles, $\theta = 0$ and $\pi$, whereas at a small $\Psi$ values, this local minimum is shifted to $\theta = \pi/2$. The orientation angles of $\theta = 0$ and $\pi$ correspond to the formation of the protrusion pattern, and $\theta = \pi/2$ can be classified as the bridge pattern. For a quantitative comparison with the experimental results, we calculated the probability distribution function of the orientation of aligned MTs at the tip \cite{beppu1,beppu2}. The probability of the orientation angle $\theta$ at the tip is given by the following equation:
\begin{equation}
P(\theta;\Psi) = \frac{\exp\bigl(- \frac{\gamma}{2D} \cos 2\theta \cos2\Psi \bigr)}{2\pi I_0\bigl(\frac{\gamma}{2D}\cos2\Psi\bigr)},
\end{equation}
where $\gamma$ is defined as $\gamma = \sum_{|\bm{r}_{mn}|<\epsilon} \bar{\gamma}$, and $I_0(\cdot)$ is the modified Bessel function of the first kind. The density distribution of MTs around the tip strongly corresponds to the theoretically predicted distribution of the preferred angle at each geometric condition ($\Delta/R = 0.50$ to 1.97) (Figure 3(d)). Intriguingly, Eq. (4) shows that the protrusion and bridge patterns occur with equal probability when $\cos2\Psi$ is zero. This implies that the transition point of two distinct patterns is at $\Psi_c=\pi/4$, yielding the following equation:
\begin{equation}
\frac{\Delta_c}{R}=2\cos \Psi_c = \sqrt{2},
\end{equation}
which is in complete agreement with the experimental findings $\Delta_c/R \approx 1.3$-1.4. Thus, we conclude that the global collective motion of confined self-propelled MTs can be controlled by geometrically defined nematic alignment at the boundary.

Notably, the protrusion pattern observed in the experiment tended to tilt slightly counterclockwise (CCW) and was not perfectly aligned from the tip in the horizontal direction ($\theta = 0$ or $\pi$). The existence of chirality in motile MTs has also been suggested \cite{yoshinaga}, and chiral MTs may contribute to the biased formation of the protrusion pattern, in either direction. Thus, it is important to determine how the chirality of the direction of motion affects the transition point from protrusion to bridge pattern. To this end, we analyzed the effect of chiral motion in numerical calculations when a chirality of $\omega = 2.0\times$10$^{-3}$ s$^{-1}$ is considered. Figure 3b shows that the CCW protrusion pattern occurred in the tilted direction, as observed in the experiment. The orientation distribution, including the effect of MT chirality, is $P(\theta;\Psi) \propto \exp\bigl[- \frac{\gamma}{2D}(\cos 2\theta \cos2\Psi - \frac{2\omega}{\gamma}\theta)\bigr]$ (see Supplemental Information). However, since the experimental value is $\omega = 2.5\times$10$^{-3}$ s$^{-1}$ with $\gamma = 1.9\times$10$^{-2}$ s$^{-1}$ and $D = 5.2\times$10$^{-3}$ s$^{-1}$ (Figure S3), the transition point $\Delta_c/R$ is estimated to be within 1.3-1.4 (Figure S4). Thus, the transition of the two patterns governed by $\Delta_c/R \approx 1.4$ still holds in the presence of slight chirality. 

A geometric rule (Eq. (5)) can account for the two distinct patterns in collective motion by limiting the collision angle guided by the boundary shape. In many biological systems, ordered cytoskeletal structures are self-organized under geometric confinement. For example, the bridge pattern found in this study closely resembles the contractile FtsZ cytoskeleton involved in the division of rod-shaped bacteria \cite{schwille} or the regulation of plant cell shape \cite{mirabet}. To explore the relevance of our findings to biological systems, we next attempted to control the collective motion of motile MTs and facilitate multiple bridge formation in a microchannel device where predesigned obstacles are attached to the boundary wall (Figure 4(a)). In a microchannel of width \SI{60}{\micro\meter}, triangle-shaped obstacles of \SI{20}{\micro\meter} in a perpendicular line were placed at regular intervals of \SI{200}{\micro\meter}. The apex angle of this obstacle, $2\Psi$, varied from $2\Psi = 0.11 \pi$ to $\pi$ (Figure 4(b)). 

We found that regularly spaced bridge formations appeared from $0.18 \pi < 2\Psi < 0.42 \pi$, whereas, under $0.54 \pi < 2\Psi < \pi$, a protrusion pattern extending in the transverse direction was frequently observed (Figure 4(c)). According to the geometric rule presented in Eq. (5), for angles less than $2 \Psi < 0.5 \pi$, a bridge connecting the protrusions is formed, and the channel becomes compartmentalized. In contrast, at angles satisfying $2 \Psi > 0.5 \pi$, a bundled protrusion is formed that leads in one direction through the channel. Hence, the geometric rule in Eq. (5) can determine the transition of confined MTs from the aligned bundled structure, parallel to the wall, to the bridge structure partitioning the compartments, with the simple design of minute obstacles along the wall of the microchannels.

\section{Conclusion}

Currently, biomolecular motors are used as a model system to explore self-organization, such as the collective motion of self-propelled cytoskeleton proteins, for nanotechnological applications. Here, we developed a fabricated PDMS device with doublet and triplet overlapping circle boundaries to reveal a geometric rule that may enable the control of the configuration of self-propelled MTs. The formation of two patterns of MTs in collective motion, that is, protrusion-like patterns and bridge formation, can be controlled by only a single geometric parameter. Intriguingly, the transition between the two patterns is determined from the nematic orientation observed in the self-propelled rod model. Moreover, this geometric rule allows us to design the assembled structure of MTs, such as by placing protruding structures in the microchannel to make a compartment wall. In addition to kinesin-driven MTs, active cytoskeletons with other molecular motors, such as myosin and dynein, are known to form ordered structures on a markedly larger scale than their size \cite{sumino,sakamoto}. As a step toward molecular actuations in nanodevices, our findings can serve as a foundation for the programmable assembly of a broad class of self-propelled cytoskeletons with a simple geometric rule.

\begin{figure*}[tbp]
\centering
\includegraphics[scale=0.78,bb=0 0 432 402]{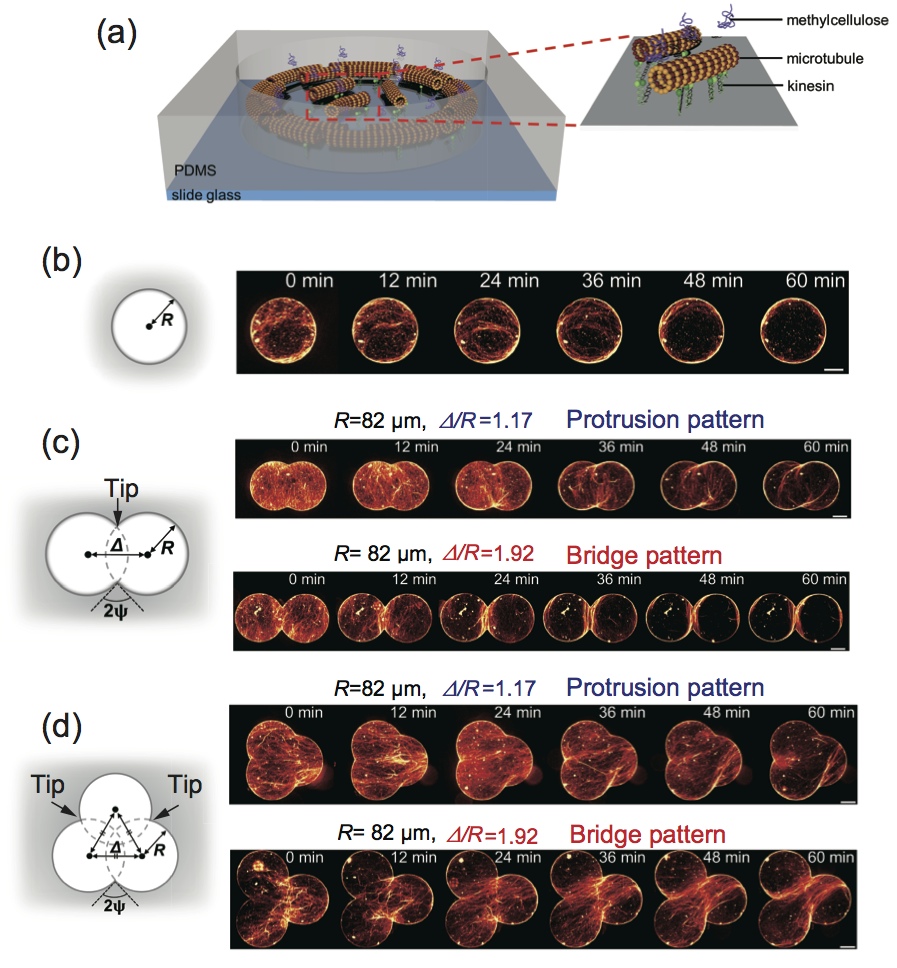}
\caption{\textbf{The collective motion of microtubules (MTs) driven by kinesin motors confined in designed microwells.} (a) Schematic illustration of the experimental setup. The kinesin molecular motors were attached to the bottom surface of the PDMS substrate. MTs are moved by the force exerted by the kinesin motor proteins. The radius of the microwell shown below is $R = \SI{82}{\micro\meter}$. (b) Time series of collective motion in MTs driven by kinesin confined in a circular microwell. (c) Time series of collective motion of MTs confined in the doublet circle microwell with (top) $\Delta = \SI{96}{micro\meter}$ and $\Delta/R = 1.17$, (bottom) $\Delta = \SI{157}{\micro\meter}$ and $\Delta/R = 1.92$. The bundled protrusions of MTs can be observed from the tip at $\Delta/R = 1.17$, while the bridge of MT bundles can be built at $\Delta/R = 1.92$. The transition of the two patterns is geometry-dependent. (d) Time series of collective motion of MTs confined in the triplet circular microwell with (top) $\Delta = \SI{96}{\micro\meter}$ and $\Delta/R = 1.17$ and (bottom) $\Delta = \SI{157}{\micro\meter}$ and $\Delta/R = 1.92$. Similar protrusions were formed from the tip at $\Delta/R = 1.17$. In contrast, these protrusions are bundled and eventually form bridges among the three tips at $\Delta/R = 1.92$.}\label{fig1}
\end{figure*}

\begin{figure*}[tbp]
\centering
\includegraphics[scale=0.78,bb=0 0 482 552]{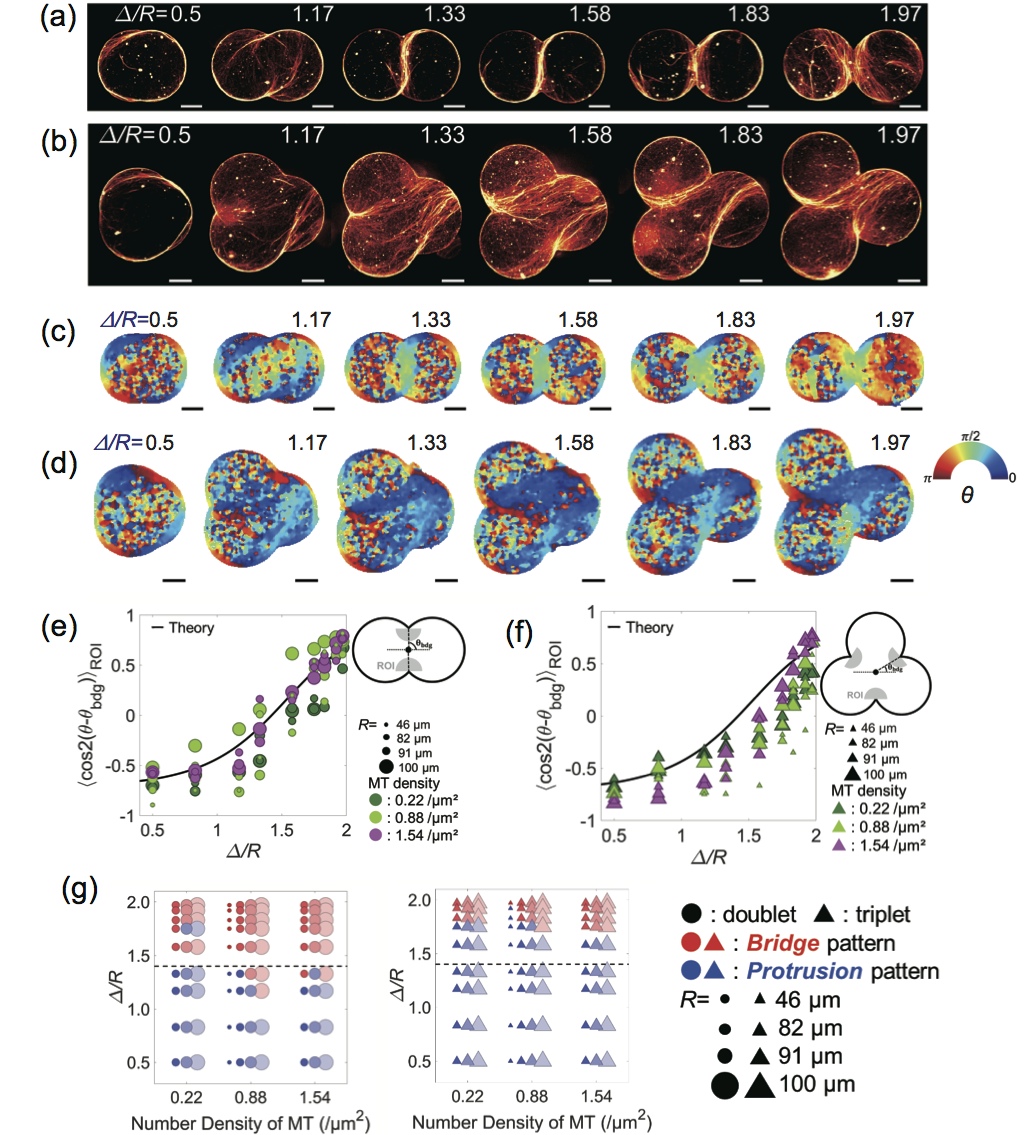}
\caption{\textbf{Geometric dependence of the protrusion and bridge patterns.} (a) Assembled patterns of self-propelled MTs at a doublet circle boundary. The distance, $\Delta$, between the centers of the two circles is varied, while the radius of the unit circle, $R=\SI{82}{\micro\meter}$, is kept constant. (b) Collective motion of MTs confined to the triplet circle boundary. As shown in (a), the distance between the centers of two adjacent circles, $\Delta$, varied, while $R=\SI{82}{\micro\meter}$ was kept constant. (c) Orientation field $\theta(\bm{r})$ of MTs at the doublet circle boundary. (d) Orientation field $\theta(\bm{r})$ of MTs at the triplet circle boundary. (e) Order parameter of the MT orientation in the doublet circle boundary at various values of $\Delta/R$. (f) Order parameter of MT orientation in the triplet circle boundary at various values of $\Delta/R$. The black solid lines in (e) and (f) are obtained from Eq. (4) with the fitting parameter $\gamma = 1.9\times$10$^{-2}$ s$^{-1}$. Inset figure in (e) and (f): ROI for order parameter calculation. (g) Phase diagram of protrusion and bridge patterns in doublet (left panel, circle) and triplet circle boundaries (right panel, triangle). The radius of the single circle of the microwell varied as follows: $R =46, 82, 91$, \SI{100}{\micro\meter}; the density of MTs is 0.22, 0.88, amd 1.54 filaments/$\mu$m$^2$. The bridge pattern is shown in red, and the protrusion pattern is shown in blue. The dotted line in (g) denotes $\Delta_c/R = \sqrt{2}$.}\label{fig2}
\end{figure*}

\begin{figure*}[tbp]
\centering
\includegraphics[scale=0.65,bb=0 0 662 402]{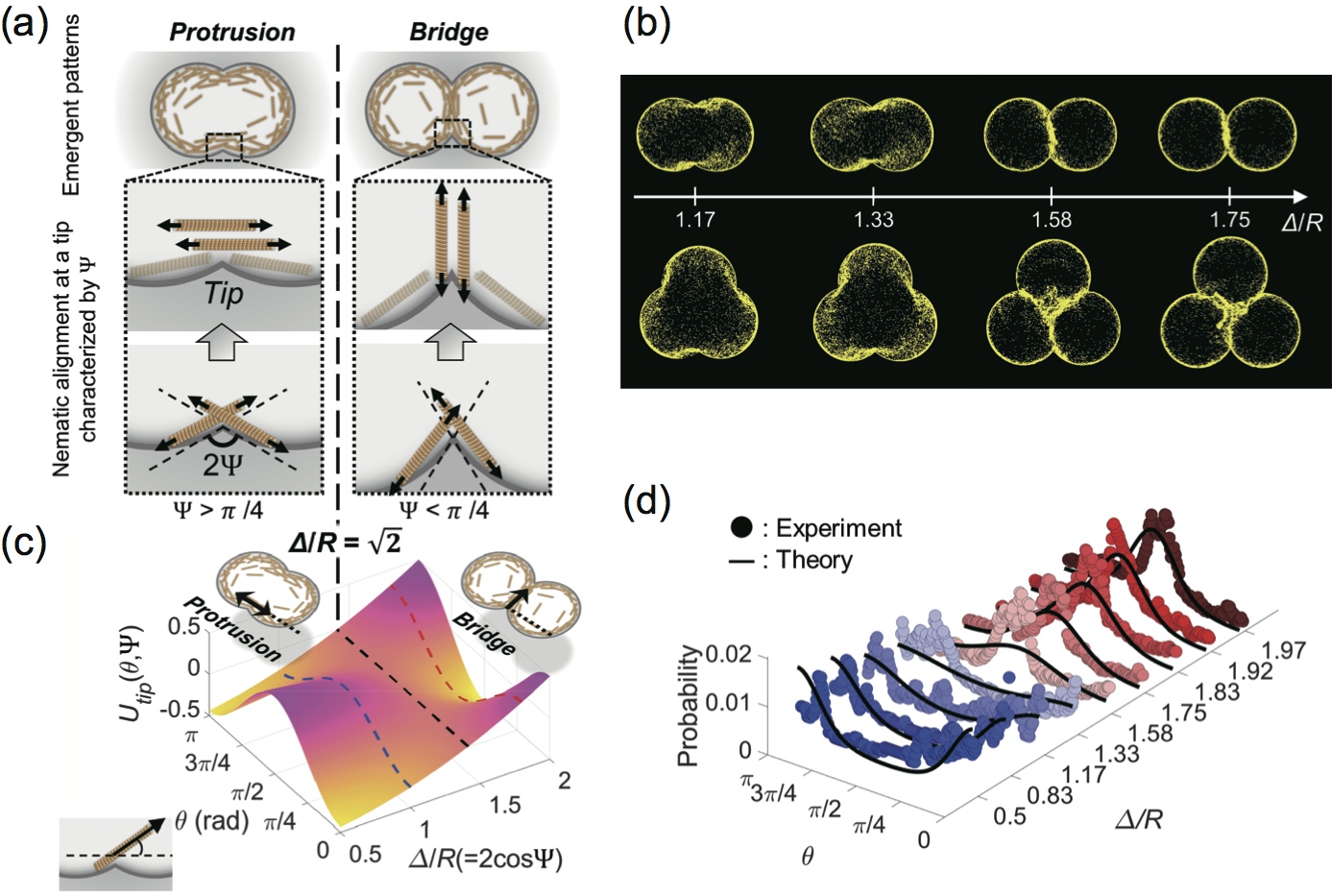}
\caption{\textbf{The mechanism of protrusion and bridge patterns analyzed by the theoretical model of self-propelled rods in confined geometries.} (a) The orientation interaction of MTs and their geometric dependence during the formation of (left) protrusion and (right) bridge patterns. The transition of these patterns depends on whether the angle of the tip, $2\Psi$, exceeds $\pi/2$. (b) Numerical simulation of self-propelled rods is performed based on an agent-based model, with the boundary of either the doublet or triplet overlapping circle. The density of the particles is 1.54 filaments/$\mu$m$^{2}$, and the other parameters are $\gamma = 5.0\times$10$^{-2}$ s$^{-1}$ and $D = 2.0\times$10$^{-3}$ s$^{-1}$, which is consistent with the experimental values. (c) Potential landscape of the geometric dependence of the orientation interaction that separates the two pattern formations. The potential of the nematic orientation interaction between agent rods was calculated for each geometrical parameter $\Delta/R$ and orientation angle $\theta$. The higher the potential in red, the lower the probability that the agent rod takes the corresponding orientation for that geometric parameter. The lower the potential (yellow), the more stable the agent rods orient themselves. (d) Comparison of theoretical calculations and experimental results for geometrical dependence of the rod orientation angle. We evaluated the number density of oriented MTs based on their fluorescence intensity near the tip. }\label{fig3}
\end{figure*}

\begin{figure*}[tbp]
\centering
\includegraphics[scale=0.78,bb=0 0 432 402]{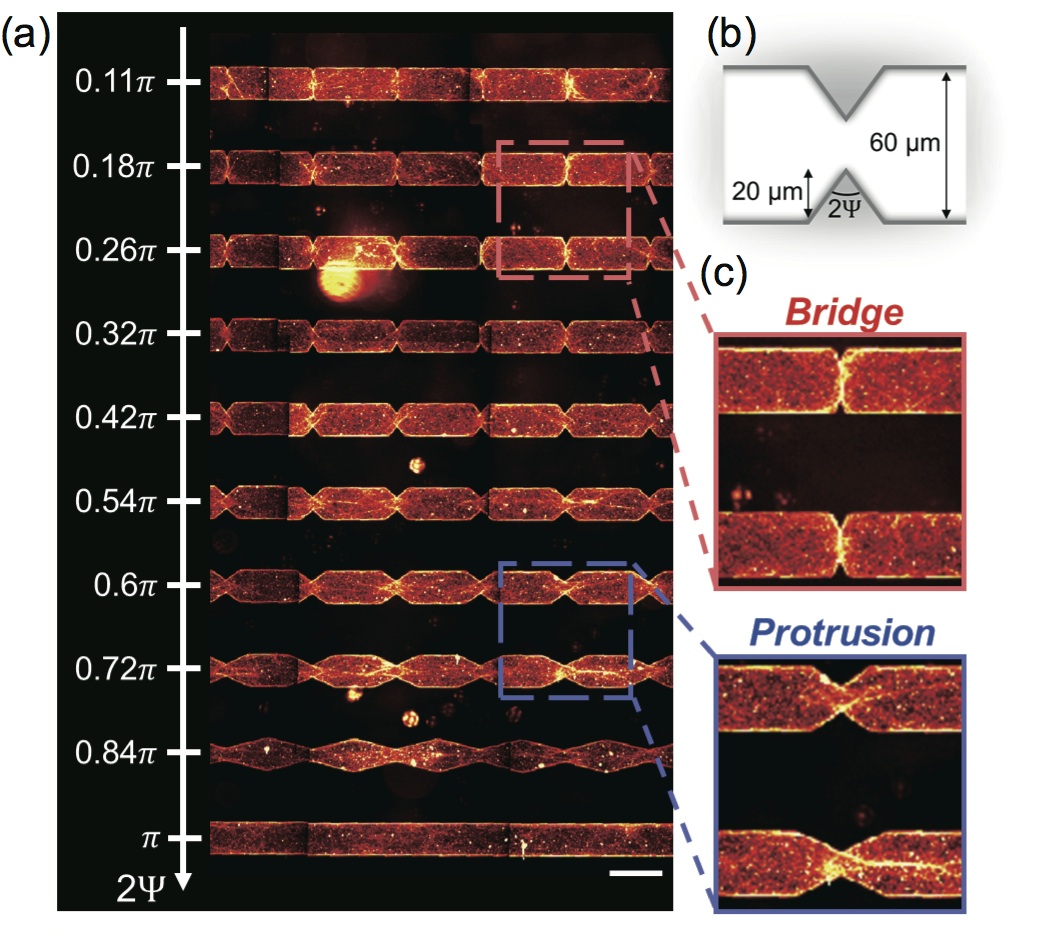}
\caption{\textbf{The partitioning of microcompartments by geometrically controlled MT bundles.} Bridge patterns were evident up to $2\psi = 0.26 \pi$, and protrusion patterning appeared along the channel from $2\psi = 0.54 \pi$ to $0.84 \pi$. The scale bar is \SI{100}{\micro\meter}.}\label{fig4}
\end{figure*}

\newpage

\begin{acknowledgement}
We thank Z. Izri for his contribution at early phase of this study. This work was supported by Grant-in-Aid for Scientific Research on Innovative Areas ``Molecular Engines" JP18H05427 (to Y.T.M.) and JP18H05423 (to K.A.), Grant-in-Aid for Scientific Research (B) JP20H01872 (to Y.T.M.), Grant-in-Aid for Challenging Research JP21K01872 (to Y.T.M.), Grant-in-Aid for Scientific Research (A) JP21H04434 (to K.A.), Grant-in-Aid for scientific research (C) JP21K04846 (to A.M.R.K.), Grant-in-Aid for Transformative Research Areas (A) JP20H05972 (to A.M.R.K.). K.B. is supported by a fellowship from Japan Society for the Promotion of Science.
\end{acknowledgement}

\section*{Author contributions statement}

Y.T.M. and A.K. designed research. S.A. conceived the experiments and analyzed data. K.B. designed microdevices and constructed theoretical models. A.M.R.K. prepared the microtubules and kinesin motors. S.A. and K.B. prepared the figures. Y.T.M. wrote the manuscript. All authors reviewed the manuscript. S.A. and K.B. are equally contributed to this work.

\bibliography{achemso-demo}

\begin{thebibliography}{45}
\bibitem{needleman} Needleman, D.; Dogic, Z. Active matter at the interface between materials science and cell biology. \textit{Nat. Rev. Materials}, \textbf{2017}, 2, 17048.

\bibitem{schwille} Zieske, K.; Schwille, P. Reconstitution of self-organizing protein gradients as spatial cues in cell-free systems. \textit{Elife}, \textbf{2014}, 3, e03949.

\bibitem{diez} Reuther, C.; Mittasch, M.; Naganathan, S.R.; Grill, S.W.; Diez, S. Highly-efficient guiding of motile microtubules on non-topographical motor patterns. \textit{Nano Lett.}, \textbf{2017}, 17, 5699-5705.

\bibitem{zoher1} Bonnemay, L.; Hostachy, S; Hoffmann, C; Gautier, J.; Gueroui, Z. Engineering spatial gradients of signaling proteins using magnetic nanoparticles. \textit{Nano Lett.}, \textbf{2013}, 13, 5147-5152.

\bibitem{zoher2} Hoffmann, C.; Mazari, E; Gosse, C; Bonnemay, L.; Hostachy, S.; Gautier, J.; Gueroui, Z. Magnetic control of protein spatial patterning to direct microtubule self-assembly. \textit{ACS Nano}, \textbf{2013}, 7, 9647-9654.

\bibitem{heuvel} van den Heuvel, M.G.L.; Dekker, C. Motor proteins at work for nanotechnology. \textit{Science}, \textbf{2007}, 313, 333-336

\bibitem{dogic1} Sanchez, T.; Chen, D.T.N.; DeCamp, S.J.; Heymann, M.; Dogic, Z. Spontaneous motion in hierarchically assembled active matter. \textit{Nature}, \textbf{2012}, 491, 431-434.

\bibitem{sumino} Sumino, Y.; Nagai, K.H.; Shitaka, Y.; Tanaka, D.; Yoshikawa, K.;Chat\'{e}, H; Oiwa, K. Large-scale vortex lattice emerging from collectively moving microtubules. \textit{Nature}, \textbf{2012}, 483, 448-452.

\bibitem{senoussi} Senoussi, A.; Kashida, S.; Voituriez, R.; Galas, J-C.; Maitra, A.; Estevez-Torres, A. Tunable corrugated patterns in an active nematic sheet. \textit{Proc. Natl. Acad. Sci. U.S.A}, \textbf{2019}, 116, 22464-22470.

\bibitem{kakugo1} Inoue, D.; Mahmot, B.; Kabir, A.M.R.; Farhana, T.I.; Tokuraku, K.; Sada, K.; Konagay, .; Kakugo, A. Depletion force induced collective motion of microtubules driven by kinesin. \textit{Nanoscale}, \textbf{2015}, 7, 18054-18061.

\bibitem{kakugo2} Keya, J.J.; Suzuki, R.; Kabir, A.M.R.; Inoue, D.; Asanuma, H.; Sada, K.; Hess, H.; Kuzuya, A.; Kakugo, A. DNA-assisted swarm control in a biomolecular motor system. \textit{Nature Commun.}, \textbf{2018}, 9, 453.

\bibitem{kakugo3} Matsuda, K.; Kabir, A.M.R.; Akamatsu, N.; Saito, A.; Ishikawa, S.; Matsuyama, T.; Ditzer, O.; Islam, M.S.; Ohya, Y.; Sada, K.; Konagaya, A.; Kuzuya, A.; Kakugo, A. Artificial smooth muscle model composed of hierarchically ordered microtubule asters mediated by DNA Origami nanostructures. \textit{Nano Lett.}, \textbf{2019}, 19, 3933-3938.

\bibitem{yoshinaga} Kim, K.; Yoshinaga, N.; Bhattacharyya, S.; Nakazawa, H.; Umetsu, D.; Teizer, W. Large-scale chirality in an active layer of microtubules and kinesin motor proteins. \textit{Soft Matt.}, \textbf{2018}, 14, 3221-3231.

\bibitem{tanida} Tanida, S.; Furuta, K.; Nishikawa,K.; Hiraiwa, T.; Kojima, H.; Oiwa, K.; Sano, M. Gliding filament system giving both global orientational order and clusters in collective motion. \textit{Phys. Rev. E}, \textbf{2020}, 101, 032607.

\bibitem{dogic3} Opathalage, A.; Norton, M.M.; Juniper, M.P.N.; Langeslay, B.; Aghvami, S.A.; Fraden, S.; Dogic, Z. Self-organized dynamics and the transition to turbulence of confined active nematics. \textit{Proc. Natl. Acad. Sci. U.S.A.}, \textbf{2019}, 116, 4788-4797.

\bibitem{wioland1} Wioland, H.; Woodhouse, F.G.; Dunkel, J.; Goldstein, R.E. Ferromagnetic and antiferromagnetic order in bacterial vortex lattices. \textit{Nature Physics}, \textbf{2016},12, 341-345.

\bibitem{beppu1} Beppu, K.; Izri, Z.; Gohya, J.; Eto, K.; Ichikawa, M.; Maeda, Y.T. Geometry-driven collective ordering of bacterial vortices. \textit{Soft Matter}, \textbf{2017}, 13, 5038-5043.

\bibitem{nishiguchi} Nishiguchi, D.; Aranson, I.S.; Snezhko, A.; Sokolov, A. Engineering bacterial vortex lattice via direct laser lithography, \textit{Nature Commun.}, \textbf{2018}, 9, 4486.

\bibitem{reinken} Reinken, H.; Nishiguchi, D.; Heidenreich, S.; Sokolov, A.; Bar, M.; Klapp, S.H.L.; Aranson, I.S. Organizing bacterial vortex lattices by periodic obstacle arrays, \textit{Commun. Phys.}, \textbf{2020}, 3, 76.

\bibitem{beppu2} Beppu, K.; Izri, Z.; Sato, T.; Yamanishi, Y.; Sumino, Y.; Maeda, Y.T. Edge current and pairing order transition in chiral bacterial vortices, \textit{Proc. Natl. Acad. Sci. USA}, \textbf{2021}, 118, e2107461118.

\bibitem{cammann} Cammann, J.; Schwarzendahl, F.J.; Ostapenko, T.; Lavrentovich, D.; Baumchen, O.; Mazza, M.G. Emergent probability fluxes in confined microbial navigation. \textit{Proc. Natl. Acad. Sci. USA}, \textbf{2021}, 118, e2024752118.

\bibitem{dogic4} Wu, K-T., Hishamunda, J.B., Chen, D.T.N., DeCamp, S.J., Chang, Y-W., Fernandez-Nieves, A., Fraden, S., Dogic, Z. Transition from turbulent to coherent flows in confined three-dimensional active fluids, \textit{Science},  \textbf{2017}, 355, eaal1979.

\bibitem{dekker} van den Heuvel, M.G.L.; Butcher C.T.; Smeets, R.M.M; Diez, S.; Dekker, C. High rectifying efficiencies of microtubule motility on kinesin-coated gold nanostructures. \textit{Nano Lett.} \textbf{2005}, 5, 1117-1122.

\bibitem{kakugo4} Inoue, D.; Gutmann, G.; Nitta, T.; Kabir, .M.R.; Konagaya, A.; Tokuraku, K.; Sada, K.; Hess, H.; Kakugo, A. Adaptation of patterns of motile filaments under dynamic boundary conditions. \textit{ACS Nano}, \textbf{2019}, 13, 12452-12460.

\bibitem{kabir} Kabir, A.M.R.; Inoue, D.;  Hamano, Y.; Mayama, H.; Sada, K.; Kakugo, A. Biomolecular motor modulates mechanical property of microtubule. \textit{Biomacromolecules}, \textbf{2014}, 15, 1797-1805.

\bibitem{izri} Izri, Z.; Garenne, D.; Noireaux, V.; Maeda Y.T. Gene expression in on-chip membrane-bound artificial cells, \textit{ACS Synth. Biol.}, \textbf{2019}, 8, 1705-1712.

\bibitem{vicsek} Vicsek, T.; Czirok, A.; Ben-Jacob, E.; Cohen, I.; Shochet, O. Novel type of phase transition in a system of self-driven particles. \textit{Phys. Rev. Lett.}, \textbf{1995}, 75, 1226.

\bibitem{chate1} Ginelli, F.; Peruani, F.; B\"{a}r, M.; Chat\'{e}, H. Large-scale collective properties of self-propelled rods. \textit{Phys. Rev. Lett.}, \textbf{2010}, 104, 184502.

\bibitem{mirabet} Mirabet, V.; Krupinski, P.; Hamant, O.; Meyerowitz, E.M.; Jonsson, H.; Boudaoud, A. The self-organization of plant microtubules inside the cell volume yields their cortical localization, stable alignment, and sensitivity to external cues. \textit{PLoS Comput Biol}, \textbf{2018}, 14, e1006011. 

\bibitem{sakamoto} Sakamoto, R.; Tanabe, M.; Hiraiwa, T.; Suzuki, K.; Ishiwata, S-i.; Maeda, Y.T.; Miyazaki, M. Tug-of-war between actomyosin-driven antagonistic forces determines the positioning symmetry in cell-sized confinement, \textit{Nat. Commun.}, \textbf{2020}, 11, 3063. 

\end{thebibliography}

\begin{thebibliography}{10}
\bibitem{castoldi} Castoldi, M.; Popov, A.V. \textit{Protein Expression and Purification}, \textbf{2003}, 32, 83-88.
\bibitem{case}  Case, R.B.; Pierce, D.W.; Nora, H.B.; Cynthia, L.H.; Vale, R.D. \textit{Cell}, \textbf{1997}, 90, 959-966.
\bibitem{peloquin} Peloquin, J.; Komarova, Y.; Borisy, G. \textit{Nature Methods}, \textbf{2005}, 2, 299-303.
\bibitem{beppu1} Beppu, K.; Izri, Z.; Gohya, J.; Eto, K.; Ichikawa, M.; Maeda, Y.T. Geometry-driven collective ordering of bacterial vortices. \textit{Soft Matter}, \textbf{2017}, 13, 5038-5043.
\bibitem{izri} Izri, Z.; Garenne, D.; Noireaux, V.; Maeda Y.T. Gene expression in on-chip membrane-bound artificial cells, \textit{ACS Synth. Biol.}, \textbf{2019}, 8, 1705-1712.
\bibitem{beppu2} Beppu, K.; Izri, Z.; Sato, T.; Yamanishi, Y.; Sumino, Y.; Maeda, Y.T. Edge current and pairing order transition in chiral bacterial vortices, \textit{Proc. Natl. Acad. Sci. USA}, \textbf{2021}, 118, e2107461118.
\bibitem{vicsek} Vicsek, T.; Czirok, A.; Ben-Jacob, E.; Cohen, I.; Shochet, O. Novel type of phase transition in a system of self-driven particles. \textit{Phys. Rev. Lett.}, \textbf{1995}, 75, 1226.
\bibitem{chate1} Ginelli, F.; Peruani, F.; B\"{a}r, M.; Chat\'{e}, H. Large-scale collective properties of self-propelled rods. \textit{Phys. Rev. Lett.}, \textbf{2010}, 104, 184502.
\bibitem{peruani} Peruani, F.; Deutsch, A.; B\"{a}r, M. A mean-field theory for self-propelled particles interacting by velocity alignment mechanisms, \textit{Euro. Phys. J.}, \textbf{2008}, 157, 111-122.
\end{thebibliography}

\newpage

\setcounter{figure}{0}
\renewcommand{\thefigure}{S\arabic{figure}}
\renewcommand{\figurename}{FIG.}
\renewcommand{\tablename}{TABLE.}
\renewcommand{\thetable}{S\arabic{table}}
\renewcommand{\refname}{References}
\renewcommand{\arraystretch}{1.2}

\section{Supporting Information}

\subsection{Materials and Methods}

\subsubsection{Protein purification}
Tubulin was purified from porcine brain using high-concentration PIPES buffer (1 M PIPES, 20 mM EGTA, 10 mM MgCl$_2$; pH adjusted to 6.8 using KOH) \cite{castoldi}. The concentration of the purified tubulin was estimated by measuring the absorbance at 280 nm and using the molar absorption coefficient=115,000 M$^{-1}$cm$^{-1}$. The kinesin used in this study was a recombinant kinesin-1 consisting of the first 573 amino acid residue of human kinesin-1 (K573) \cite{case}. The concentration of the purified kinesin was estimated from sodium dodecyl sulphate-polyacrylamide gel electrophoresis (SDS-PAGE). Bovine serum albumin (BSA) solutions of various concentrations were used as the standard. Using the image processing software ‘ImageJ’, a standard curve was prepared from the concentration of BSA solutions and intensity of corresponding BSA bands in the SDS-PAGE result. The concentration of kinesin was estimated from kinesin band intensity and the standard curve obtained using the BSA solutions. High-molarity PIPES buffer (HMPB) and 2×BRB80 buffer were prepared using PIPES from Sigma, by adjusting the pH using KOH. EGTA and MgCl$_2$ were purchased from Dojindo and Fujifilm Wako, Japan respectively.

\begin{figure*}[b]
\centering
\includegraphics[scale=0.7,bb=0 0 702 402]{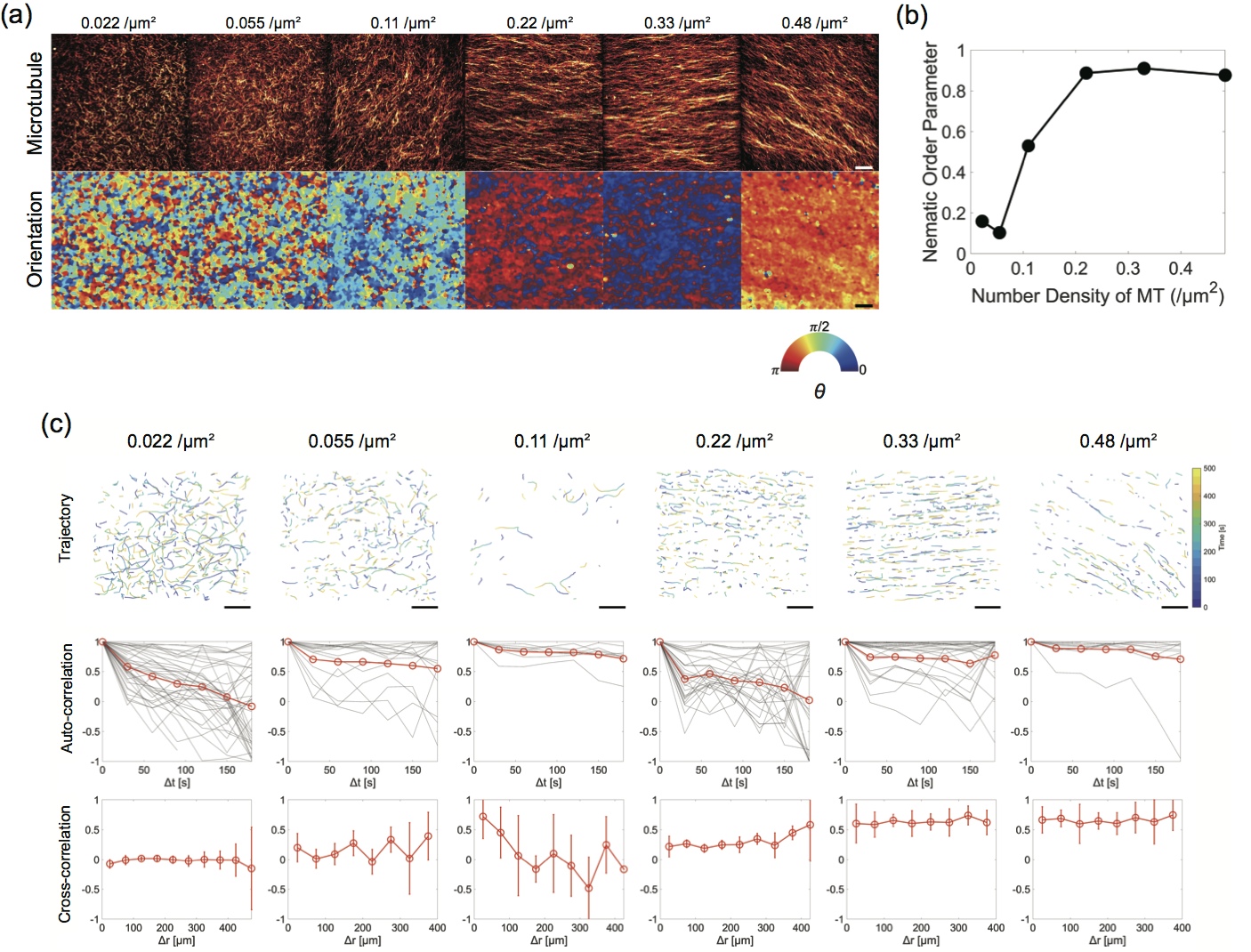}
\caption{Collective motion of kinesin-driven microtubules at various microtubule densities. (a) Top: Fluorescence images of motile microtubules showing collective motion. Bottom: orientation field of aligned microtubules calculated from corresponding fluorescence images. Scale bar: \SI{50}{\micro\meter}. (b) Nematic order parameters at various microtubule densities. (c) Velocity correlation function of kinesin-driven microtubules at various microtubule densities. (Top) Typical trajectory of motile microtubules at various microtubule densities. The color code represents acquisition time. (Middle) Autocorrelation function of velocity of motile microtubules in time at various microtubule densities. (Bottom) Cross-correlation function of velocity of motile microtubules at distance $r$ and various microtubule densities. Scale bar: \SI{50}{\micro\meter}.}
\end{figure*}

\subsubsection{Microtubule polymerization}
Rhodamine-labeled tubulin was prepared using 5/6-carboxytetramethylrhodamine succinimidyl ester (TAMRA-SE; Invitrogen) following a standard protocol \cite{peloquin}. The labeling ratio was 1.0 as determined by measuring the absorbance of the protein at 280 nm and that of tetramethyl-rhodamine at 555 nm. Rhodamine-labeled microtubules (MTs) were prepared by polymerizing a mixture of rhodamine tubulin (RT) and non-labeled tubulin (WT) for 30 min at \SI{37}{\degreeCelsius} (RT/WT = 4:1; final tubulin concentration = 56 $\mu$M). The prepared MTs were kept in paclitaxel buffer (50 $\mu$M in BRB80 buffer) overnight at \SI{25}{\degreeCelsius}. The microtubule solution was diluted with wash buffer (80 mM PIPES, 1 mM EGTA, 2 mM MgCl$_{2}$, 0.5 mg mL$^{-1}$ casein, 1 mM DTT, 10 $\mu$M paclitaxel, and 1\% DMSO; pH 6.8) before use.

\subsubsection{Gliding assay}
Flow cells were built with a double-sided tape. Surface cleaning was applied on flow cell for 12 min in a plasma cleaner (Harrick). We then injected \SI{10}{\micro\liter} of Casein solution at 0.5 mg/mL into the flow cell. After short incubation, \SI{10}{\micro\liter} of kinesin solution (80mM PIPES, 1 mM EGTA, 1mM MgCl$_{2}$, 700 nM kinesin; pH 6.8) was applied and incubated for 5 min. The flow cell was washed with 10 $\mu$L of wash buffer (80 mM PIPES, 1 mM EGTA, 1mM DTT, 0.5 mg/mL Casein, 4.5 mg/mL D-Glucose, 50 U/mL Glucose oxidase, 50 U/mL Catalase, 2 mM MgCl$_2$, 50 $\mu$M Taxol; pH 6.8). Next, \SI{10}{\micro\liter} of microtubule solution of prescribed concentration was applied and incubated 10 min, followed by washing with \SI{10}{\micro\liter} of wash buffer with 0.3 wt\% methylcellulose. Then, \SI{10}{\micro\liter} of ATP buffer (80 mM PIPES, 1 mM EGTA, 1mM DTT, 0.5 mg/mL-Casein, 4.5 mg/mL D-Glucose, 50 U/mL Glucose oxidase, 50 U/mL Catalase, 2 mM MgCl$_2$, 50 $\mu$M Taxol, 5 mM ATP, 0.3 wt\% methylcellulose; pH 6.8) was applied to the flow cell. All the experiments were performed at room temperature.

In the gliding assay experiment with a glass substrate, we examined the ordered streaming of kinesin-driven microtubules on the glass surface and captured a fluorescence image of rhodamine-labeled microtubules 60 min after the addition of ATP buffer. The pattern of aligned microtubules, such as long-range streaming, can be observed at a microtubule density $\rho \geq 0.22$ filaments/$\mu$m$^{2}$ (Figure S1(a)). The orientation field $\theta(\bm{r})$ at position $\bm{r}$ is calculated from the fluorescence image, and we found highly oriented microtubule patterns similar to the microtubule density $\rho \geq 0.22$ filaments/$\mu$m$^{2}$. To characterize this density-dependent pattern formation, we used the nematic order parameter $S$ in the orientation field $\theta(\bm{r})$. The nematic order parameter $S$ is defined as $\langle e^{2i \theta(\bm{r})}\rangle$, where $\langle \cdot \rangle$ is the ensemble average over the region of interest. As the density of the microtubule increases, the nematic order parameter $S$ gradually increases and reaches close to 1 (highly aligned state) at approximately $\rho = 0.22$ filaments/$\mu$m$^{2}$ (Figure S1(b)). We chose the microtubule density $0.88$ filaments/$\mu$m$^{2}$, which is sufficiently dense to induce the collective motion of motile microtubules, for typical experiments described in the main text (Figure 1 and 2).

In addition, we examined the velocity $\bm{v}(\bm{r},t)$ of a single motile microtubule and its directionality at various densities (Figure S1(c), top). The autocorrelation function of velocity $C_a(\Delta t)= \langle \bm{v}(\bm{r},t)\cdot\bm{v}(\bm{r},t+\Delta t)\rangle_t$ shows a similar relaxation curve irrespective of the density of the microtubules (Figure S1(c), middle). This result indicates that the directionality of motile microtubules is persistent and not affected by the density of microtubules. In contrast, the cross-correlation function of the velocity of different microtubules, defined as $C_c(\Delta r) = \langle \bm{v}(\bm{r},t)\cdot\bm{v}(\bm{r}+\Delta \bm{r},t)\rangle_{\bm{r}}$, shows a monotonic increase as the density of the microtubules increases (Figure S1(c), bottom). This result is consistent with the monotonic increase of the nematic order parameter S in a density-dependent manner because the ordered streaming of microtubules reflects the aligned self-propelling of single microtubules.

\begin{figure*}[tb]
\centering
\includegraphics[scale=0.65,bb=0 0 702 402]{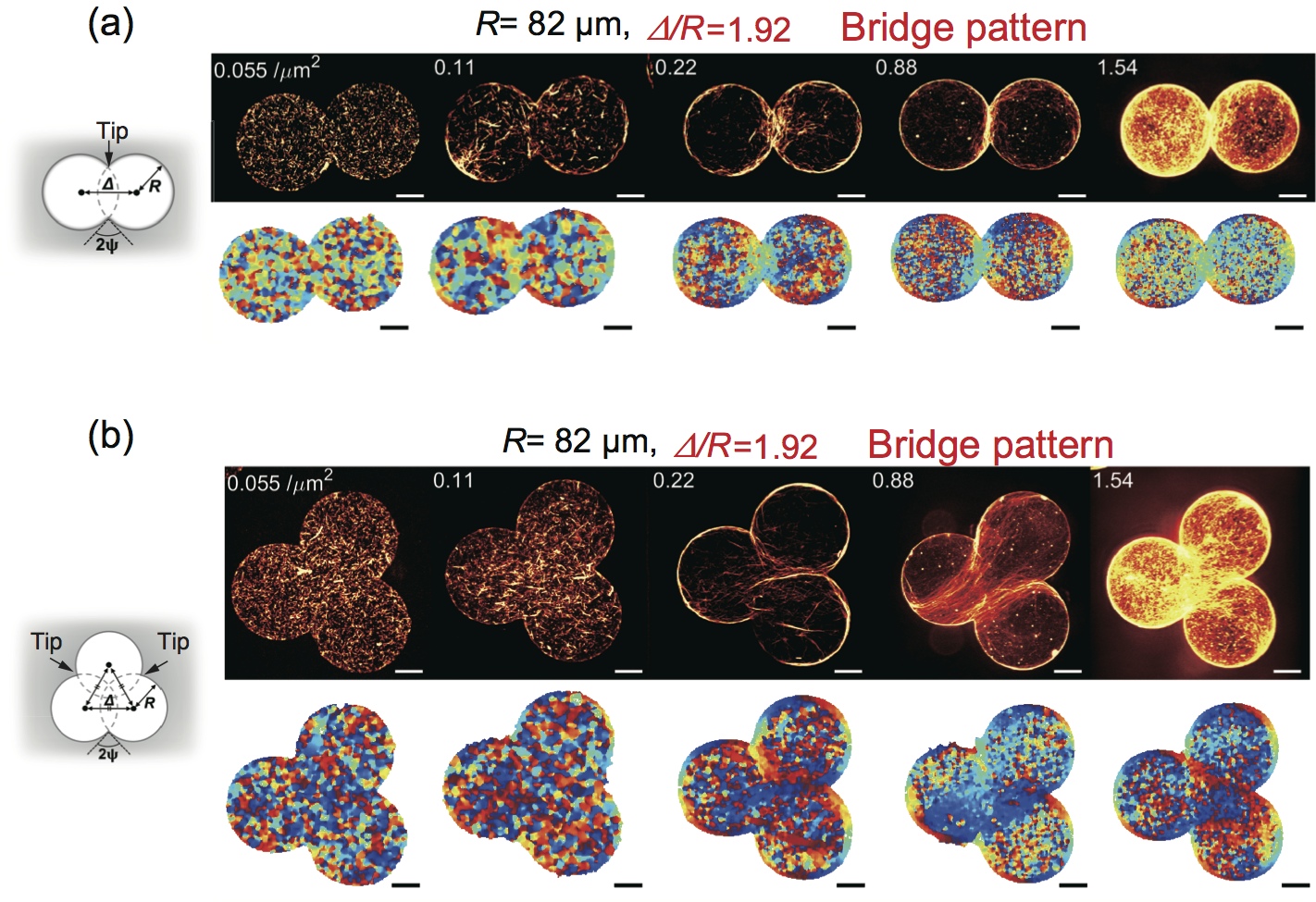}
\caption{Collective motion of kinesin-driven microtubules confined in designed microwells at various microtubule densities. (a) Top: Fluorescence images of motile microtubules with collective motion at the doublet circle boundary. Bottom: Orientation field of aligned microtubules calculated from corresponding fluorescence images. (b) Top: Fluorescence images of the collective motion of motile microtubules in triplet circle boundary. Bottom: Orientation field of aligned microtubules calculated from corresponding fluorescence images. Scale bar: \SI{50}{\micro\meter}. }
\end{figure*}

\subsubsection{Microfabrication}
The collective motion of microtubules and kinesin molecular motors was observed while they were spatially constrained in poly-dimethyl siloxane (PDMS, Sylgard 184, Dow Corning) microwells. A template of SU-8 photoresist (Microchem, SU8 3025) was prepared by conventional photolithography, and PDMS was poured into this template and cured to create microwells. The patterned PDMS sheet was transferred to a glass substrate, and the PDMS surface was treated with plasma cleaning (Harrick, PDC-32G) to prepare kinesin-bound microwells \cite{beppu1,izri,beppu2}. The glass attached to the PDMS sheet was used as the substrate for the flow cell. The kinesin motor protein was adsorbed on the PDMS surface, and the microtubules were enclosed in the flow cell chamber. An example of the typical collective dynamics of the confined microtubules is shown in Figure S2. The collective motion of motile microtubules becomes evident at a microtubule density of $\rho \geq 0.22$ filaments/$\mu$m$^{2}$, which is consistent with that presented in Figure S1. We set the microtubule density at 0.88 filaments/$\mu$m$^{2}$ for experiments described in the main text, unless otherwise specified.

\subsubsection{Optical microscopy and image analysis}
Microtubules were labeled with fluorescent dyes, and all images were acquired using a confocal microscope (IX83 inverted microscope from Olympus, CSU-X1 confocal scanning unit from Yokogawa Electric Co. Ltd., and iXon-Ultra EM-CCD camera from Andor Technologies). The temperature was maintained at \SI{24}{\degreeCelsius}. In order to analyze the orientation distribution of microtubules in collective motion (Figure 2 in the main text), the fluorescence images were analyzed using orientation J in image J software. In addition, in the analysis of the fluorescence density distribution (Figure 3 in the main text), the obtained images were analyzed using the Image Processing Toolbox of MATLAB software (MathWorks).

\subsection{Theoretical section}
\subsubsection{Numerical simulation}
We describe the collective motions of the self-propelled rods under boundaries of doublet and triplet circles with radius $R$. This numerical simulation refers to our previous study \cite{beppu2}. We define the polar coordinate $(r, \varphi)$ in two-dimensional plane and the origin of the polar coordinate is set at the center of the circle the rod belongs to among two (or three) circles for the doublet (or triplet) circular boundary. We consider self-propelled rods $m$, with a position $\bm{r}_{m}(t)=(x_{m}(t), y_{m}(t))=r_m(\cos\varphi_m, \sin\varphi_m)$ and an orientation $\bm{d}(\theta_{m}) = (\cos\theta_m, \sin\theta_m)$, moves at a constant speed $v_0$ under a geometric constraint. The time evolution of the position of the rod is given by 
\begin{equation}
\dot{\bm{r}}_m = \bm{v}(\theta_m(t)) + \frac{2\kappa}{l^2} \sum_{r_{mn} <\varepsilon}(\bm{r}_m - \bm{r}_n)\exp\Biggl[-\biggl(\frac{r_{mn}}{l}\biggr)^2\Biggr] -\kappa_b \frac{\bm{r}_{m}}{r_{m}} \Theta(r_{m}-R),
\end{equation}
where $r_{mn} = |\bm{r}_m - \bm{r}_n|$ denotes the distance between two rods $m$ and $n$, and $\varepsilon$ is the range of nematic orientation interaction. To prevent rods from overlapping at the lateral wall, we incorporate soft repulsion with the coefficient $\kappa$ into the minimal Vicsek-style model. $l$ is the length scale of soft repulsive interaction. $\kappa_b$ represents the coefficient of the repulsive force to confine the rod in the multiplet circular boundary. $\Theta(x)$ represents Heaviside step function defined as 
\begin{equation}
\Theta(x) =\left\{\begin{array}{ll}1, & x>0 \\ 0, & x \leq 0\end{array}\right.
\end{equation}

In the calculation of the orientation dynamics of self-propelled rods, we consider the nematic interaction among rods that works for neighbor rods within the range of $r_{mn} <\varepsilon$. Moreover, the nematic interaction with the wall is considered to impose a reorientation at the boundary wall. Because motile MTs align almost perfectly parallel to the wall in the experiment, we simply set the strong nematic alignment that immediately makes the rod on the wall align in its tangential direction, allowing us to precisely test the effect of a collision angle at a tip on the global alignment pattern. Summing up all alignment effects, the following equation gives the orientation dynamics of the rod \cite{vicsek,chate1},
\begin{equation}
\dot{\theta}_m = \omega - \bar{\gamma}\sum_{r_{mn}<\varepsilon} \sin2(\theta_m - \theta_n) + \eta_m,
\end{equation}
where $\omega$ is the angular velocity in chiral motion, $\bar{\gamma}$ is the coefficient of nematic alignment of the adjacent rods, and $\eta(t)_m$ is angular fluctuation ($\langle \eta_m \rangle =0$, $\langle \eta_m (t)\eta_n (t')\rangle=2D\delta_{mn}\delta(t-t')$).

We perform the simulations using the following parameters: $v_0 = \SI[per-mode=symbol]{0.3}{\micro\meter\per\second}$, $\bar{\gamma} = \SI[per-mode=symbol]{0.05}{\per\second}$, $D = \SI[per-mode=symbol]{0.002}{\per\second}$, $\omega = \SI[per-mode=symbol]{0.002}{\per\second}$, $\varepsilon = \SI{0.3}{\micro\meter}$, $\kappa = \SI[per-mode=symbol]{0.2}{\micro\meter^2\per\second}$, $l= \SI{0.1}{\micro\meter}$, $\kappa_b = \SI[per-mode=symbol]{10}{\micro\meter\per\second}$, $R = \SI{40}{\micro\meter}$. The equations were integrated using the Euler method, with $\delta t = 0.01$. The integration was conducted up to $T = 800$. We set the number density of rods to be 1.54 rods/$\mu$m$^{2}$, the same as the experimental highest value. Fig. 3(b) in the main text shows the array of snapshots of self-propelled rod vectors confined in the doublet and triplet circular boundary at the end point ($T = 800$). For comparison with the experimental results regarding order parameter of the bridge formation, we analyze the coarse-grained orientation filed with the mesh size $\Delta x = \SI{1.6}{\micro\meter}$. In the same way as the experimental analysis, we calculate the order parameter in the semicircular regions around tips with radius $R/6$ by averaging over $T = 701$ to 800 with the sampling rate $\Delta t = 1$ (Figure S4(a)). The transition point from protrusion patterns to bridge ones appears to be around $\Delta/R = 1.4$, consistent with the experimental result.

\subsubsection{Derivation of Eq. (4) in the main text}
We consider self-propelled rods that can interact through a potential $U$ of the nematic alignment. The state of rod $m$ at time $t$ is represented by two variables: its coordinate $\bm{r}_m(t)$ and orientational angle $\theta_m(t)$. The particles align their orientation through $\partial U(\bm{r}_m, \theta_m)/\partial \theta$, and the coefficient of nematic alignment is given by $\bar{\gamma}$. The rods move at a constant speed of $v_0$. Hence, the evolution of $\bm{r}_m(t)$ and $\theta_m(t)$ for two-dimensional coordinates is given by the self-propelled rod model as follows:
\begin{equation}\label{position}
\dot{\bm{r}}_m = v_0 \bm{e}(\theta_m)
\end{equation}
where $\bm{e}(\theta_m)$ is the unit vector of velocity defined as $\bm{e}(\theta_m)=(\cos\theta_m, \sin\theta_m)$. 
\begin{equation}\label{orientation}
\dot{\theta}_m = - \bar{\gamma} \frac{\partial U}{\partial \theta}+ \eta_m(t) 
\end{equation}
where $\eta_m(t)$ represents the random noise in the angular direction of the rod, and its correlation satisfies $\langle\eta_m(t)\rangle=0$, $\langle\eta_m(t)\eta_n(t\prime) \rangle =2D\delta_{mn}\delta(t-t\prime)$, where $\delta_{mn}$ is the Kronecker delta, $\delta(t)$ is the Dirac delta function, and $D$ is the angular diffusion constant. The alignment of velocity vector is based on nematic interaction, and thus, the potential $U$ is given by the following equation
\begin{equation}
U(\bm{r}_m, \theta_m) = - \sum_{r_{mn}<\varepsilon} \cos2(\theta_m - \theta_n) \end{equation}
where $r_{mn} = |\bm{r}_m - \bm{r}_n|$, and $\varepsilon$ is the effective radius of the nematic interaction. 

As shown in Figure 3(a) presented in the main text, geometric confinement induces the transition from the protrusion to bridge pattern as the value of $\Delta/R$ is increased. We consider the interaction of motile rods from the left or right circle in a doublet microwell defined by the geometric constant $\Psi$ ($\Delta/R = 2\cos\Psi$). In the tip, where the two circles intersect, bacteria swimming in the left and right wells interact and become oriented. For instance, if a motile rod moves in the counterclockwise (CCW) direction (or clockwise [CW] direction) along the wall in the left microwell, the heading angle is $\pi/2 - \Psi$ (or $-\pi/2 - \Psi$) at the tip. In contrast, a motile rod moving in the CW direction (or CCW direction) in the right microwell has a heading angle of $\pi/2 + \Psi$ (or $-\pi/2 + \Psi$) at the tip. The orientational dynamics in Eq. \eqref{orientation} can be rewritten as follows:
\begin{equation}\label{orientation2}
\dot{\theta}_m = \frac{\gamma}{4}\bigl(\sin2(\theta_m - \pi/2 + \Psi) + \sin2(\theta_m + \pi/2 + \Psi) + \sin2(\theta_m - \pi/2 - \Psi) + \sin2(\theta_m + \pi/2 - \Psi)\bigr) + \eta_m(t).
\end{equation}
where $\gamma = \bar{\gamma} \sum_{r_{mn}<\varepsilon}$. Next, the dynamics of the rod orientation is reduced, as shown by the following equation:
\begin{equation}\label{orientation2}
\dot{\theta}_m = - \gamma \sin2\theta_m\cos2\Psi + \eta_m(t),
\end{equation}
where $U(\theta;\Psi) = \frac{1}{2}\cos2\theta_m\cos2\Psi$.

By considering the mean field approximation \cite{peruani}, we can derive the Fokker-Planck equation for the probability distribution of rod orientation, $P(\theta,t;\Psi)$, as follows:
\begin{equation}\label{FP}
\frac{\partial P}{\partial t} = D \frac{\partial^2 P}{\partial \theta^2} + \gamma \frac{\partial}{\partial \theta} \Bigl[(\sin2\theta \cos 2\Psi) P \Bigr].
\end{equation}

Because our aim is to control the static pattern of confined self-propelled rods, the left-hand side of Eq. \eqref{FP} is set at $\frac{\partial P}{\partial t} = 0$. By solving Eq. \eqref{FP} under this steady state condition, we can find the probability distribution $P (\theta;\Psi)$ appearing in Eq. (4) of the main text as follows:
\begin{equation}\label{soln_fp}
P(\theta;\Psi) = \frac{\exp\bigl(- \frac{\gamma}{2D} \cos 2\theta \cos2\Psi \bigr)}{2\pi I_0(\frac{\gamma}{2D}\cos2\Psi)},
\end{equation}
where $I_0(\cdot)$ is a modified Bessel function of the first kind.

\subsubsection{The effect of chiral motion of self-propelled rods}

We analyzed the effect of chiral motion of kinesin-driven microtubules for confined collective motion as well as the resulting geometric rule by considering chiral rotation in orientational dynamics. Given that the motile microtubule has intrinsic chirality in its motion, the angular rotation is constant at $\omega$. By taking a minimal self-propelled rod model, the orientational dynamics with chiral rotation are as follows:
\begin{equation}\label{orientation3}
\dot{\theta}_m = - \bar{\gamma} \frac{\partial U}{\partial \theta} + \omega + \eta_m(t).
\end{equation}
The same calculation performed in the previous section can be applied to Eq. \eqref{orientation3} with the additional term of $\omega$. The alignment of chiral microtubules at the tip of the confined microwells is given as follows:
\begin{equation}\label{orientation4}
\dot{\theta}_m = - \gamma \sin2\theta_m\cos2\Psi + \omega + \eta_m(t),
\end{equation}
and the Fokker-Planck equation is given as follows:
\begin{equation}\label{FP2}
\frac{\partial P}{\partial t} = D \frac{\partial^2 P}{\partial \theta^2} + \gamma \frac{\partial}{\partial \theta} \biggl[\Bigl(\sin2\theta \cos 2\Psi - \frac{\omega}{\gamma}\Bigr) P \biggr].
\end{equation}

By solving Eq. \eqref{FP2} under steady state conditions ($\frac{\partial P}{\partial t} = 0$), the probability distribution, $P(\theta;\Psi)$, for rod orientation at the tip is extended to the following form:
\begin{equation}\label{soln_fp_chiral2}
P(\theta;\Psi) = \frac{\exp\Bigl(- \frac{\gamma}{2D} \bigl[\cos 2\theta \cos2\Psi - \frac{2\omega}{\gamma} \theta \bigr]\Bigr)}{2\pi I_0(\frac{\gamma}{2D}\cos2\Psi)}.
\end{equation}

The solid line in Figure S4(b) shows the transition point from the protrusion to bridge pattern, calculated using Eq. \eqref{soln_fp_chiral2}. The experimental value of $\omega/\gamma$ is approximately 0.1, which has only a negligible effect on the protrusion-bridge pattern transition of self-propelled microtubules.

\begin{figure*}[tb]
\centering
\includegraphics[scale=0.78,bb=0 0 362 402]{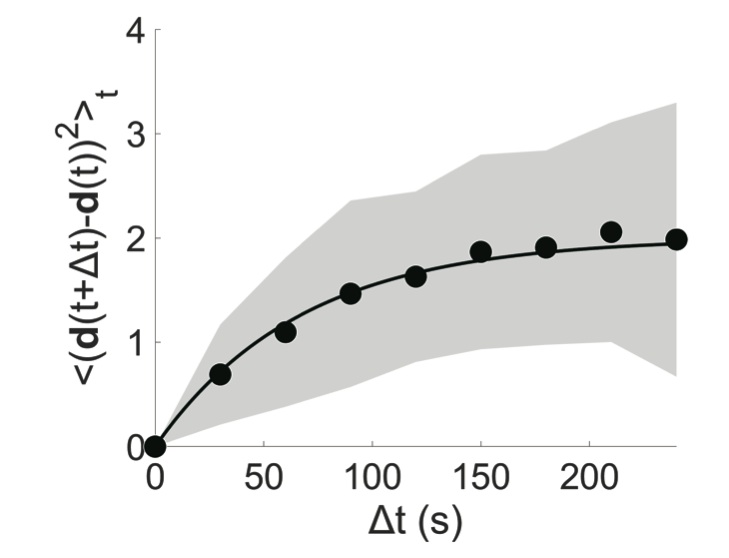}
\caption{ The mean square angular displacement (MSAD) of the orientation angle of motile microtubules is plotted with time. The slope of this MSAD curve is fitted with $2(1-\exp[-D \delta t])$, where $D$ reflects the orientation fluctuation.}
\end{figure*}

\begin{figure*}[tb]
\centering
\includegraphics[scale=0.73,bb=0 0 632 402]{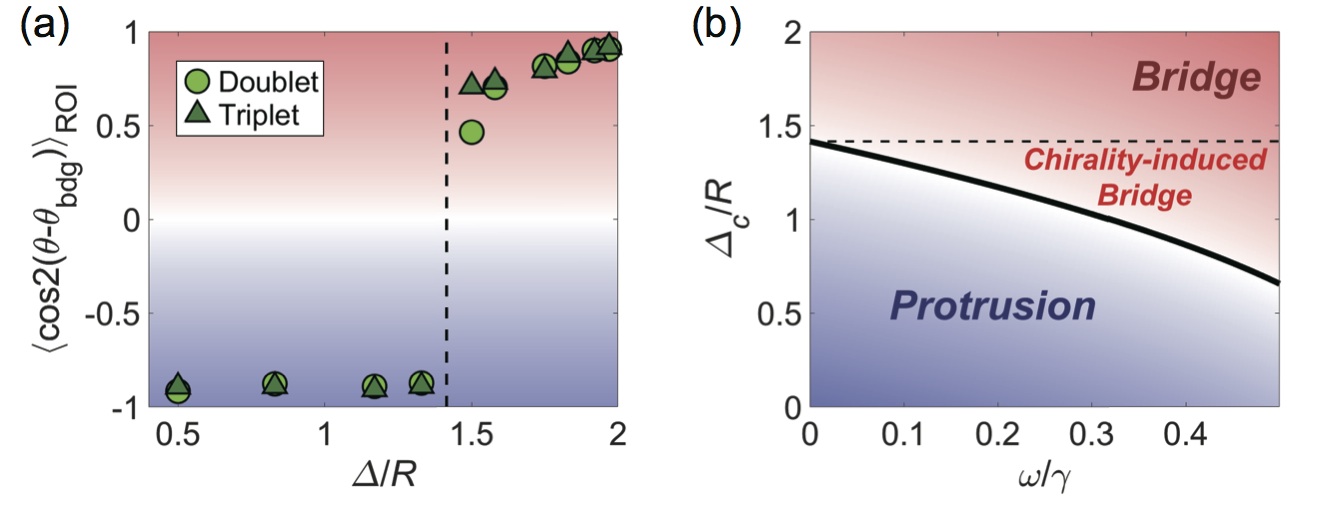}
\caption{Theoretical analysis of self-propelled rods in numerical simulations. The dotted line indicates the geometric rule $\Delta_c/R=\sqrt{2}$. (a) Order parameter of collective motion of self-propelled rods in a confined geometry of doublet and triplet circle boundaries. (b) Phase diagram of protrusion and bridge patterns under the effect of chirality in motion. The phase diagram is shown with the geometric parameter $\Delta/R$ and angular speed of the chiral self-propelled rod.}
\end{figure*}

\end{document}